\documentstyle{spacekp} 

\begin{opening}
\title{DISTRIBUTION OF GAMMA-RAY BURSTS \\ IN HALO NEUTRON STAR-COMET MODELS}

\author{M.G. Higgins}
\author{R.N. Henriksen}
\institute{Queen's University at Kingston \\ Kingston, Ontario, Canada}
\date{}
\end{opening}

\runningtitle{DISTRIBUTION OF GRBS IN HALO NEUTRON STAR-COMET MODELS}
\runningauthor{M.G. HIGGINS AND R.N. HENRIKSEN}

\begin{document}

\begin{abstract}

The motions of comets and neutron stars have been integrated over five 
billion years in the Galactic potential to determine a gamma-ray burst  
distribution, presuming that bursts are the result of interactions between
these two families of objects. The comets originate in two distinct 
populations - one from ejection by stars in the Galactic disk, and the other
from ejection by stars in globular clusters. No choice of the free parameters
resulted in agreement with both the isotropy data and the $\log(N>F)$-$\log(F)$
data.

\end{abstract}


\section{Introduction}

There have been many models over the years which try to identify gamma-ray
bursts (GRBs) with
neutron stars (NSs) in an extended Halo around the Galaxy which interact in some
way with passing comets (e.g. Pineault and Duquet (1993) or Mitrofanov and
Sagdeev (1990)). The details of the environment which would 
allow such a scenario involving interstellar comets have not been worked out in 
detail.

Over the
age of the solar system, it has been estimated that approximately $10^{12}$
comets have escaped the Sun's gravitational well (Stern (1990)). If 
other solar systems behave in a similar manner, there should be a large
number of free comets in globular clusters (GCs), which, after escaping the
clusters by scattering, could seed an extended Halo.

This work evolved a population of $10^6$ NSs born in the disk. It also evolved 
$10^7$ comets
ejected from GCs, and assumed a population of disk comets which follow the
mass density of the disk. Using these results, a sample of GRBs was created.

\section{The Model of Galactic Potential}

The model used was taken from Paczynski (1990), with some modifications. It 
assumes three contributions to the total potential: a disk, a spheroid, and
a spherical dark matter halo.
The values used for the constants in his model, relating to the oblateness of
the spheroid ($a_1$ and $b_1$) and disk ($a_2$ and $b_2$), are:
$a_1 = 0$, $b_1 = 0.277$ kpc, $a_2 = 3.7$ kpc, and $b_2 = 0.2$ kpc. A dark 
matter halo with constant density inside a radius $r_{\rm c} = 8$ kpc,
and density falling off like $r^{-2.25}$ outside $r_{\rm c}$, replaced 
Paczynski's halo.
The masses used are:
M$_1 = 1.12 \times 10^{10} {\rm M}_\odot$ (the spheroid mass), 
M$_2 = 8.07 \times 10^{10} {\rm M}_\odot$ (the disk mass),
M$_{\rm c} = 1.75 \times 10^{10} {\rm M}_\odot$ (the dark halo mass inside 
$r_{\rm c}$).

Other components contributing to the Galactic potential, such as satellite
galaxies or the globular cluster population, were ignored for simplicity.
Over the course of the simulation, roughly 10 \% of the comets would
be affected by such objects; this is not expected to change the results to
any great degree. In addition, it is assumed that the Galactic dark matter
does not interact with the comets to any appreciable degree. This assumption
would fail if the dark matter is mainly molecular clouds, for example.

\section{The Three Populations}
\label{sec-comets}

The GC population comets were ejected
from the GC stellar systems with small relative speeds. 
A small fraction of these 
comets escaped from the GC without
scattering; however, a much larger fraction were scattered by stars in the
GC (for simplicity, all scattering stars were taken to have a mass M$_{\odot}$),
and were ejected from the GC with speeds approximately 10-20 km/s.
140 typical GCs were used in the simulation. The GC population was found
by assuming a radial distribution which falls off like $r^{-0.85}$ (Kulessa
and Lynden-Bell (1992)),
and a velocity distribution with gaussian form, with average value
$v_{0} = 220 - 2 r_{\rm kpc}$ km/s, 
and standard deviation
$\sigma = 120 - r_{\rm kpc}$ km/s.
Here, $r_{\rm kpc}$ is the radial distance from the centre of the 
Galaxy in
kpc, allowed to range from 6-110 kpc. Each GC was taken to have mass $10^5$ 
M$_\odot$, and the Plummer 
model, with a radial distance scale of 1 pc, was used for the gravitational 
potential.

The disk population of comets was taken to follow the disk
component of the total density.
The total number of disk comets was set to the total number of GC comets
multiplied by the ratio of the disk mass to the total mass of GCs, and a factor
$f$ (one of the free parameters in the work), which is a ratio of the
efficiencies of comet ejection in the disk to comet ejection in the GCs.

The NSs were started in the disk, with a position distribution which follows
the observed pulsar distribution, taken from Johnston (1994). 
The initial speeds considered ranged from 350 km/s to 2000 km/s in a
frame of reference rotating around the Galactic centre.

\section{Estimating the Interaction Probability}

The probability of interaction between the NS and comet populations
was taken to be (e.g. Pineault and Poisson (1989)) 
$P = \frac{A n_{\rm ns} n_{\rm c}}{v_{\rm rel}}$.
Here, $A$ is a constant relating to the cross-section for collisions, 
$n_{\rm ns}$ is the NS number density, normalised to unity at the
maximum value,
$n_{\rm c}$ is the normalised total comet number density 
(disk comets plus GC comets), and
$v_{\rm rel}$ is the average relative speed of the two populations at the point
in question, determined in the simulations. This includes only gravitational
focussing in the collisions.

The burst frequency in this model can be roughly estimated, but since
much of the physics is unknown, assigning values to the constants should be 
taken as 
tentative at best. Very approximately, $\frac{dN_{\rm bursts}}{dt} = N_{\rm c} 
\tilde{n}_{\rm NS} \sigma_{\rm int} v_{\rm rel}$, where $N_{\rm c}$ is the 
total number of comets,  
$\tilde{n}_{\rm NS}$ is an average number density of NSs, and $\sigma_{\rm int}$
is the interaction cross-section.

The total number of comets is given by 
$N_{\rm c} = N_{\rm sGC} N_{\rm cGC} + N_{\rm sd}
 N_{\rm cd}$, where $N_{\rm sGC}$ is the total number of GC 
stars ($\simeq 10^{7}$), $N_{\rm sd}$ is the total number of disk stars
($\simeq 10^{10}$), $N_{\rm cd}$ is the number of comets produced by a 
disk
star, and $N_{\rm cGC}$ is the number of comets produced by a GC star; 
using the efficiency ratio $f$ from section~\ref{sec-comets}, $N_{\rm cGC}
= N_{\rm cd}/f$.

As a rough estimate, we can use the number of comets estimated to have been
ejected from our solar system as an approximation to $N_{\rm cd}$. Assuming a 
total population of
$10^9$ NSs, most inside a sphere of radius 40 kpc, relative speeds
on the order of $10^3$ km/s, and that the cross-section is just the square of the
light-cylinder radius for a NS spinning with an angular velocity of 0.1 rad/s,
the number of bursts per day is roughly 
$\frac{dN}{dt} \simeq 4 {\rm bursts/day} (\frac{10^{-6}}{f})$.

\begin{figure}
\vspace{5cm}
\caption{Projected sky map of 5000 sample GRBs for NS birth velocities of
700 km/s, $f = 0$}
   
\end{figure}

\section{Simulated Gamma-Ray Burst Distribution}

Figure 1 shows a sky map of 5000 sample GRB positions, using a NS birth velocity
of 700 km/s and efficiency ratio $f = 0$ (no disk comets). The anisotropy 
toward the Galactic 
disk is obvious, even without including the effect of the disk comet
population. The clustering is evident
with as few as 585 bursts (as in the BATSE 2B catalog). Including a disk 
population
rapidly (for any $f > 10^{-5}$) makes clumping toward the Galactic plane 
extremely efficient.
Changing the birth velocities of NSs has little effect,
due to the clumped comet population.
Note that increasing $f$ beyond $10^{-5}$ not only causes clumping toward the
disk, but also decreases the number of bursts/day to numbers much smaller
than are observed. The requirement of such a discrepancy 
in the ejection efficiencies is another strike against this model.

The flux of a burst was calculated by letting the total energy $E$ of a burst
follow a gaussian distribution in logarithmic space, with the average
burst energy set so that a burst with this energy, located at 100 kpc, would
have a flux equal to the ``instrumental'' minimum observable flux. A standard
deviation of half an order of magnitude was used. This is a somewhat ad-hoc
value; the true distribution in $E$ likely depends on distributions of
NS magnetic field strength and spin rate, which are unknown for the burster
population. The results are not very sensitive
to this spread as along as it is less than two orders of magnitude, where it
becomes comparable to the spread in flux due to the range in distances of the
bursts.

The relation $\log (N>F) \propto -3/2 \log F$ holds for an isotropic distribution
of standard candles, where $\log (N>F)$ is the number of bursts 
with flux greater than $F$.
The simulation data are roughly consistent with this distribution for large 
values of flux,
though a slope of -1.44 is the best fit. The flattening of the slope observed
in the BATSE data is also seen in the simulation data.

\section{Conclusions}

These results suggest that models of GRBs as comet-NS collisions in an 
extended Halo are untenable for a wide range of free parameters. While  
simulation data, with a somewhat unphysical choice for the ejection efficiency
of GC stars, roughly meet the distribution in flux of real GRBs, no values
of the free parameters allow a fit to the isotropy data. However, this
work does assume that the NSs are ejected from the Galactic disk, and are not
relics from the initial formation of the Galaxy. Choosing an appropriate
distribution of relic NSs may allow a model in agreement with the observations.
In addition, replacing the spherical dark matter halo with an oblate one might
reduce the clumping toward the Galactic centre.

{}

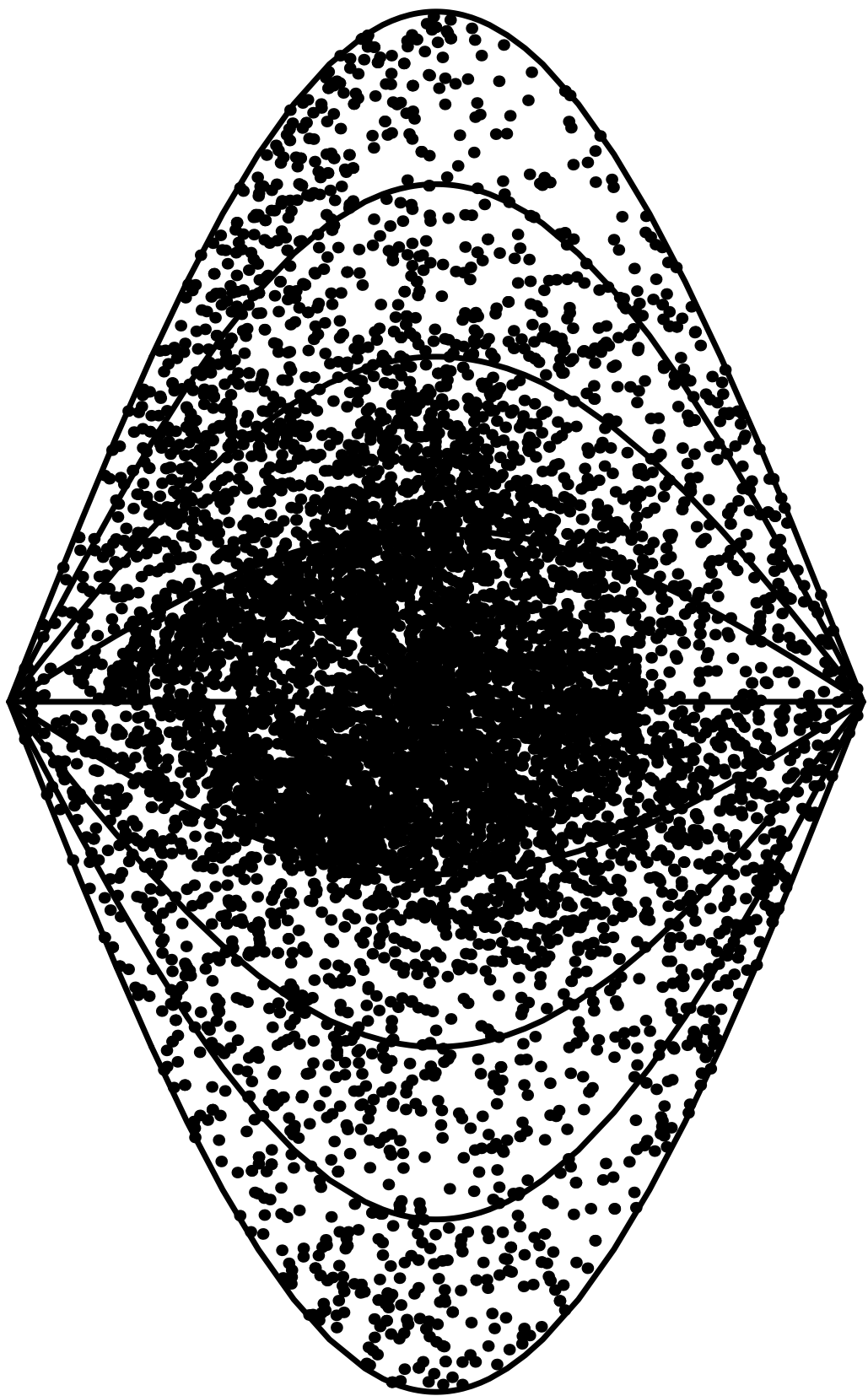
\end{document}